\documentclass[apj]{emulateapj} 

\shorttitle{The OH megamaser emission from IRAS~17208--0014}
\shortauthors{Momjian et al.}

\begin{document}

\title{Sensitive VLBI Studies of the OH Megamaser Emission from IRAS~17208--0014}

\author{E. Momjian\altaffilmark{1}}
\email{emomjian@naic.edu}

\author{J. D. Romney\altaffilmark{2}}
\email{jromney@nrao.edu}

\author{C. L. Carilli\altaffilmark{2}}
\email{ccarilli@nrao.edu}

\author{T. H. Troland\altaffilmark{3}}
\email{troland@pa.uky.edu}

\altaffiltext{1}{National Astronomy and Ionosphere Center, Arecibo Observatory, HC 3 Box
53995, Arecibo, PR 00612.}
\altaffiltext{2}{National Radio Astronomy  Observatory, P O Box O, Socorro, NM 87801.}
\altaffiltext{3}{University of Kentucky, Department of Physics and Astronomy, Lexington,
KY 40506.}

\begin{abstract}

We present phase-referenced VLBI results on the radio continuum and the OH 18~cm megamaser
emission
from the Ultra-Luminous Infrared Galaxy, IRAS~17208--0014. The observations were carried
out at 1599~MHz using the Very Long Baseline Array, the phased VLA, and the Green Bank
Telescope.
The highest resolution radio continuum results show several compact sources with
brightness temperatures on the order of $10^{6}$~K. These sources are more likely to be
clustered supernova remnants and/or luminous radio supernovae. However, the agreement
between the number of observed and expected compact sources above the 5~$\sigma$ level
supports the possibility that each one of the compact sources could be dominated by a
recently detonated luminous radio supernova. The continuum results suggest that there is
no radio-loud AGN in the nuclear region of this galaxy.
The OH 18~cm megamaser emission in IRAS~17208--0014 is detected at various angular
resolutions. It has an extent of $170 \times 110$~pc, and is mostly localized in two
regions separated by 61~pc. The structure and dynamics of the maser emission seem to be
consistent with a clumpy, rotating, ring-like geometry, with the two dominant maser
regions marking the tangential points of the proposed rotating-ring distribution.
Assuming Keplerian motion for the rotating maser ring, the enclosed dynamical mass and
the mass density within a radius of 30.5~pc, are about {$3 \times 10^7~({\rm
sin}^{-2}i)~M{_\odot}$}, and $281~({\rm sin}^{-2}~i)~M{_\odot}~{\rm pc}^{-3}$,
respectively.

\end{abstract}


\section{INTRODUCTION}
At bolometric luminosities greater than $10^{11}~L{_\odot}$, infrared galaxies
become the most numerous objects in the local universe ($z \leq 0.3$) \citep
{SM96}.
The trigger for the intense infrared emission appears to be the strong
interaction or merger of molecular gas-rich spirals.
Galaxies with the highest
infrared luminosities ($L{_{\rm IR}}(8-1000~{\mu}m) \geq 10^{12} L{_\odot}$),
known as Ultra-Luminous Infrared Galaxies (ULIRGs), appear to be advanced merger
systems powered by a nuclear starburst and/or an AGN. These may represent an
important stage in the formation of quasi-stellar
objects \citep{SAN88}.

The extreme merging conditions in ULIRGs could, in some cases, induce
nuclear OH 18~cm masing several orders of magnitude stronger than for typical Galactic
masers,
i.e., megamasers ($10^6$ times), or even gigamasers ($10^9$ times) \citep{B89,B92}.
Recent surveys by \citet{DG02} report an
OH megamaser (OHM) detection rate of up to 33\% for ULIRGs at $z>0.1$.

Sensitive Very Long Baseline Interferometry (VLBI) observations provide the most direct
means to study OHM and radio continuum emission in the innermost, dust-obscured, regions
of merging systems. While radio continuum studies play a key role to directly image and
determine the nature of the nuclear power sources (AGN and/or starburst) in these
galaxies, the OHM emission studies 
can enhance our understanding of the kinematics on parsec scales in their nuclear
regions, and be used as probes of AGNs or intense nuclear starbursts.

In the present paper, we report VLBI observations of the 18 cm radio continuum and OHM
emission from the galaxy IRAS~17208--0014 at $z=0.04281$. This object is one of the few
strong OHM emitters at $z < 0.1$ that can be imaged at milliarcsecond (mas) resolution,
and contribute to a better understanding of the nature, extent, and dynamics of
extragalactic OHM emission.

This galaxy, which is classified as a ULIRG, has an infrared luminosity of $L_{\rm
IR}= 2.5 \times 10^{12} L{_\odot}$, as defined in \citet {GOL95}.
We adopt a distance of 171~Mpc,
assuming ${H_\circ=75}$~km~s$^{-1}$~Mpc$^{-1}$. At this distance, 1~mas
corresponds to 0.76~pc.

Optical images of IRAS~17208--0014 at 6550 {\AA} show two tidal tails from a
merger \citep {MM90,MUR96}.
Near-IR images show a very disturbed morphology and an extended, but single,
nucleus, suggesting a very advanced merger \citep {ZL93,MUR96}.
Infrared observations suggest that this galaxy represents the
extreme of starburst-dominated sources of this type \citep {SOI00}, and
an observational proof that a collision of galaxies can lead to a mass
distribution similar to elliptical galaxies \citep {ZL93}.

The radio continuum emission of this galaxy at milliarcsecond (mas) resolution was
first detected by the VLBI observations of \citet {MOM03} at 21~cm. These observations
revealed that the radio emission is dominated by an extreme nuclear starburst,
with no indication of a compact high-brightness temperature source (i.e., a radio-loud
AGN).

IRAS~17208--0014 exhibits OHM activity at 1665, 1667, and 1720~MHz \citep{MAR89}. The
strongest emission is in the 1667~MHz line, with a luminosity of $L_{\rm
OH}=1054~L_{\odot}$ \citep{MAR89}.
Very Large Array (VLA) and single-dish Nan\c{c}ay observations showed the existence of two
separate velocity structures with similar widths for both the 1667 and 1665~MHz OH main
lines \citep {MAR89}. However, the 18~cm VLBI observations of \citet{DLLS99} did
not detect the 1665~MHz line, nor the continuum emission, and recovered only 20\% of the
1667~MHz line.

\section{OBSERVATIONS AND DATA REDUCTION}

The present observations were carried out at 1599~MHz on 2003 May~25,
using three NRAO~\footnote{The National
Radio Astronomy Observatory is a facility of the National Science Foundation
operated under cooperative agreement by Associated Universities, Inc.}
facilities: the Very Long Baseline Array (VLBA), the
phased VLA, and the Green Bank Telescope. The bandwidth of the
observations was 16 MHz, in each of the right- and left-hand circular polarizations, with
the central frequency set appropriately for the 1667~MHz transition of the OH 18~cm
line at a heliocentric velocity $cz=12790$~km~s$^{-1}$ \citep{MAR89,MOM03}.
The data were two-bit sampled, and correlated at the VLBA correlator in Socorro, New
Mexico, with 512-point spectral resolution per baseband channel, and 2~s correlator
integration
time. The total observing time was 10~hr.

Data reduction and analysis were performed using the Astronomical Image
Processing System (AIPS) and the Astronomical Information Processing System
(AIPS++).

Along with the target source IRAS~17208--0014, the compact source J1730+0024 was
observed as a phase reference. A cycle consisting of 180~s on
the target source and 60~s on the reference source was adapted. The source J1743--0350
was used for amplitude and bandpass calibration.

After applying a priori flagging and manually excising integrations
affected by interference, we performed amplitude calibration using
the measurements of the antenna gain and system temperature of each station,
and bandpass calibration. The flux density calibration of the data set is believed
to be accurate at the level of 5\% \citep{WU05}.

A continuum data set was generated by averaging the OH-emission-free spectral channels.
In the continuum data set, the phase reference source J1730+0024, was self-calibrated and
imaged in an iterative cycle. The self-calibration solutions for J1730+0024 were
applied to both the continuum and line data of the target source. 

The continuum and line data sets of IRAS~17208--0014 were then deconvolved and imaged at
various spatial resolutions, with a grid weighting intermediate between pure natural and
pure uniform (${\rm ROBUST}= 0$ in AIPS task IMAGR). The continuum images were obtained
using the multi-resolution CLEAN algorithm in IMAGR, which has been shown to be superior
compared to the standard CLEAN or Maximum Entropy Method in imaging a combination of very
weak extended and compact emission regions \citep{MOM03}.

\section{RESULTS}

Fig.~1 is the continuum image of the central region in IRAS~17208--0014 at 1599~MHz and
$17.6 \times 6.1$~mas ($13.4 \times 4.6$~pc, P.A. =$-4^{\circ}$) resolution, and
represents the bright emission region to the west of center in Fig.~2 of \citet{MOM03}.
This region is populated by 10 compact sources with peak flux densities $>~5\sigma$
(155~$\mu$Jy~beam$^{-1}$), but $<~7\sigma$ (217~$\mu$Jy~beam$^{-1}$).
Most of these sources were reproduced with pure natural weighting (ROBUST=5), but not with
pure uniform weighting (ROBUST=$-$5). This is largely due to the much higher noise in the
uniformly weighted image, which is a factor of 6 worse than natural weighting, and almost
a factor of 4 worse than the intermediate weighting (ROBUST=0).
Table~1 lists the Gaussian fitting parameters of the compact continuum sources seen in
Fig.~1 derived using the AIPS task JMFIT. The positions (col. [2]) are relative to the
first source in the list. Column (3) lists the surface brightnesses of these sources, and
column (4) their integrated flux densities. Column (5) gives the deconvolved sizes of the
Gaussian components at FWHM, and column (6) gives the position angles. For several of
these components, the nominal deconvolution sizes are listed as given by JMFIT. However,
for some components, no nominal sizes were obtainable for the minor axes of the fitted
Gaussians, hence, the maximum sizes of the minor axes are reported as upper
limits. Moreover, because source number 10 is unresolved, we are only able to quote an
upper limit for its overall size. The corresponding brightness temperatures of these
compact sources are on the order of $10^{6}$~K.

\begin{figure}
\epsscale{1}
\plotone{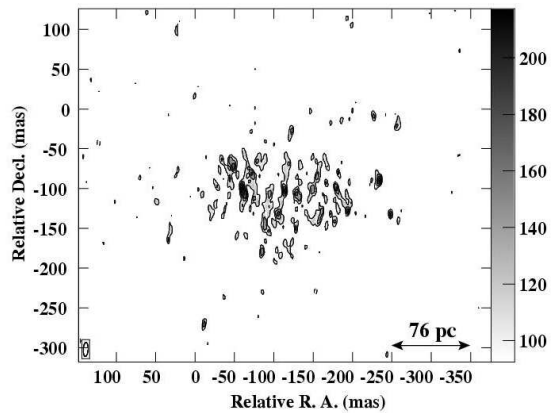}
\caption{Continuum image of the central region in IRAS~17208--0014 at 1599~MHz. The
restoring beam size is 
$17.6 \times 6.1$~mas at position angle $-4^{\circ}$. The peak flux density is
217~$\mu$Jy~beam$^{-1}$, and the contour levels are at $-$3, 3, 4, 5 and 6 times the rms
noise level, which is 31~$\mu$Jy~beam$^{-1}$. The gray-scale range is indicated by the
wedge at the right side of the image.
The reference position (0, 0) is $\alpha \rm{(J2000)}=17^{\rm h} 23^{\rm m}
21\rlap{.}^{\rm s} 9648$,
$\delta \rm{(J2000)}=-00^{\circ} 17' 00\rlap{.}'' 819$.
For comparison purposes, this origin has been chosen to be the same as for the continuum
figures of \citet{MOM03}.}
\end{figure}

Fig.~2 is a total intensity (moment 0) image of the OH 18~cm main lines (both 1665 and
1667~MHz) at 25~mas resolution. The image was obtained by blanking areas where the flux
density of the OH spectral-line image cube is below 4 $\sigma$, or 3.6~mJy~beam$^{-1}$.
Fig.~2 also shows the Hanning-smoothed spectra of the OH 18~cm main lines averaged over
various regions of the OHM emission.

OH emission from IRAS~17208--0014 is spread over a broad range of velocities. The strong,
narrow feature seen in all profiles of Fig.~2 (other than for Region 4) almost certainly
arises from the 1667 MHz transition. This conclusion seems very likely for two reasons.
Firstly, this relatively narrow feature lies close to the systemic velocity of the galaxy
if the feature arises from the 1667 MHz transition (see Section 2). Secondly, this narrow
feature is always accompanied by a weaker emission feature offset by about
370~km~s$^{-1}$, the velocity equivalent to the difference in frequency between the 1665
and 1667 MHz lines at the redshift of IRAS~17208--0014.

The overall extent of the megamser emission is $170 \times 110$~pc on the plane of the
sky, and seems to be dominated by two regions separated by 80~mas (61~pc) at position
angle 83$^{\circ}$. We designate these as the eastern and western components (Regions 1
and 5 in Fig.~2). These two regions seem to be connected by weaker maser emission
regions, interrupted in the middle by what seems to be a region devoid of maser emission.
In addition to the two main maser regions, a weak extension toward the south, and an even
weaker one toward the north are detected. The measured velocity difference between the
eastern and the western regions is $137 \pm 3$~km~s$^{-1}$.

Table~2 summarizes the physical characteristics of the dominant OH peaks seen in the
spectra of Fig.~2 for both the 1665 and the 1667 MHz transitions. The line properties
were obtained by fitting Gaussians to the non-Hanning-smoothed OH spectra obtained in the
regions specified in Fig.~2 using the tool IMAGEPROFILEFITTER in AIPS++. Column (1) lists
the region number as designated in Fig.~2. The velocities (col. [2]) refer to the center
velocities of the 1667~MHz line peaks in each region. Column (3) lists their full widths
at half maximum, and  column (4) is their peak flux density values. Columns (5)--(7), are
similar to columns (2)--(4), but for the 1665~MHz line peaks. Column (8) is the peak flux
density ratio of the 1667~MHz line to the 1665~MHz line for each region. For the spectrum
of region Region~6, which shows two distinct emission peaks for the 1667~MHz line, two
Gaussian components are listed. 
No Gaussian fitting parameters are reported for the spectrum seen in Region~4, because of
its low signal-to-noise ratio and the apparent blending of the two mainlines. 

\begin{figure}
\epsscale{1.1}
\plotone{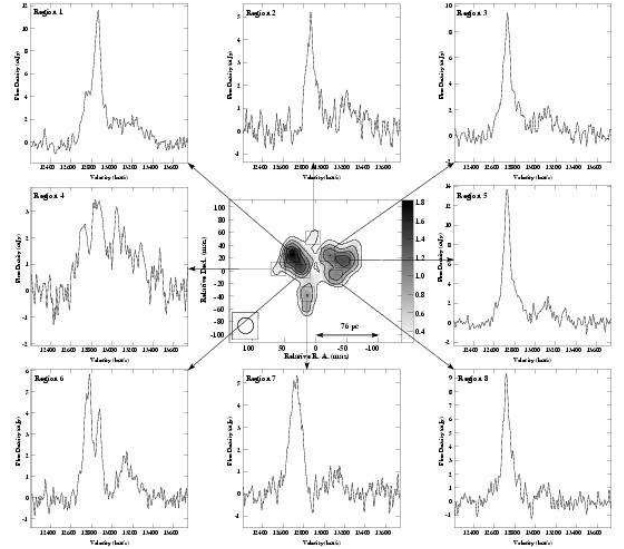}
\caption{Gray-scale and contour total intensity (moment 0) image of the OH 18~cm main
lines, and Hanning-smoothed OH spectra obtained at various locations of the OHM emission.
The restoring beam size is $25 \times 25$~mas. The contour levels are at 0.3, 0.6, 0.9,
\ldots, 1.8~Jy~beam$^{-1}$~km~s$^{-1}$, and the gray scale range is indicated by the step
wedge at the right-hand side of the image in units of Jy~beam$^{-1}$~km~s$^{-1}$. The
reference position (0, 0) is $\alpha \rm{(J2000)}=17^{\rm h} 23^{\rm m} 21\rlap{.}^{\rm
s}9533$, $\delta \rm{(J2000)}=-00^{\circ} 17' 00\rlap{.}''983$. The rms noise level of
the total intensity is 0.13~Jy~beam$^{-1}$~km~s$^{-1}$. The velocity scale of the spectra
is referenced to the 1667~MHz line, and the spectral features between 13000 and
13400~km~s$^{-1}$ are from the 1665~MHz line \citep{MAR89}. The effective velocity
resolution of the spectra is 12.3~km~s$^{-1}$}.
\end{figure}

Fig.~3 is a Hanning-smoothed spectrum of the OH 18~cm main lines integrated over the whole
OH emission region seen in Fig.~2. The velocity scale is referenced to the 1667~MHz line.
The weak emission between 13000 and 13400~km~s$^{-1}$ is from the 1665~MHz main line
\citep{MAR89}.
The arrow in this figure points to the expected velocity of the 1665~MHz line that
corresponds to the main peak of the 1667~MHz line. While the shape of the spectral
profile seen in Fig.~3 is similar to that obtained at Nan\c{c}ay and with the VLA
\citep{MAR89}, the VLBI array recovers only $\sim$80\% of the single-dish and VLA flux
densities.

\begin{figure}
\epsscale{1}
\plotone{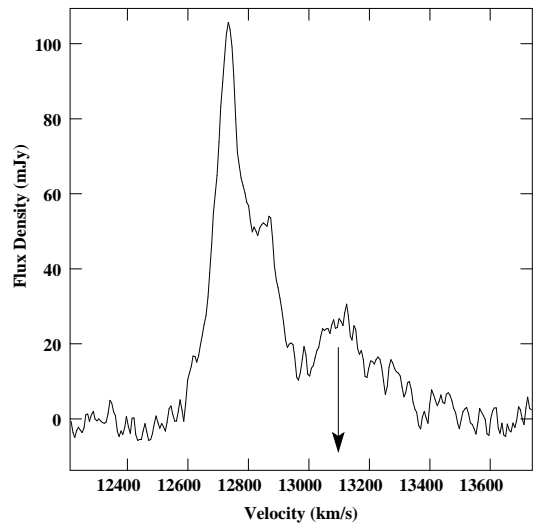}
\caption{Hanning-smoothed OH 18~cm spectrum integrated over the complete OH emission
region as seen in our VLBI observations. The velocity scale is referenced to the 1667~MHz
line. The effective velocity resolution is 12.3~km~s$^{-1}$. The arrow points to the
expected velocity of the 1665~MHz line that corresponds to the main peak of the 1667~MHz
line. At the redshift of IRAS~17208--0014, the offset between these two lines is
368~km~s$^{-1}$.}
\end{figure}

Fig.~4 displays gray-scale total intensity (moment 0) images of the 1667~MHz main line at
three different angular resolutions with continuum contours superposed. For
comparison purposes, the same origin is used for the coordinates as by \citet{MOM03} for
their continuum images. Fig.~4{\it a} is at 50~mas resolution, Fig.~4{\it b} at 25~mas
resolution, and Fig.~4{\it c} at the full resolution of the VLBI array, which is $17.6
\times 6.1$~mas resolution (P. A.$=-4^{\circ}$). These images were obtained by
restricting the spectral channel range to that of the 1667~MHz main line, and blanking
the OH image cubes at the 4~$\sigma$ level, where $\sigma$ is 1.8, 0.9, and
0.5~mJy~beam$^{-1}$, for the 50~mas, 25~mas, and the full resolution images,
respectively. In all these images, the OH emission from the eastern region appears to be
stronger than that from the west. The lower resolution images show the eastern region to
be associated with stronger continuum emission than the western. Moreover, the continuum
emission seems to have a larger extent than that of the OHM, and its peak is located 
north-east of the megamaser regions. At the full resolution, we find no correlation
between the brightest continuum and maser emission spots. The derived lower limit on the
maser amplification factor is $\sim 130$.

\begin{figure}
\epsscale{1}
\plotone{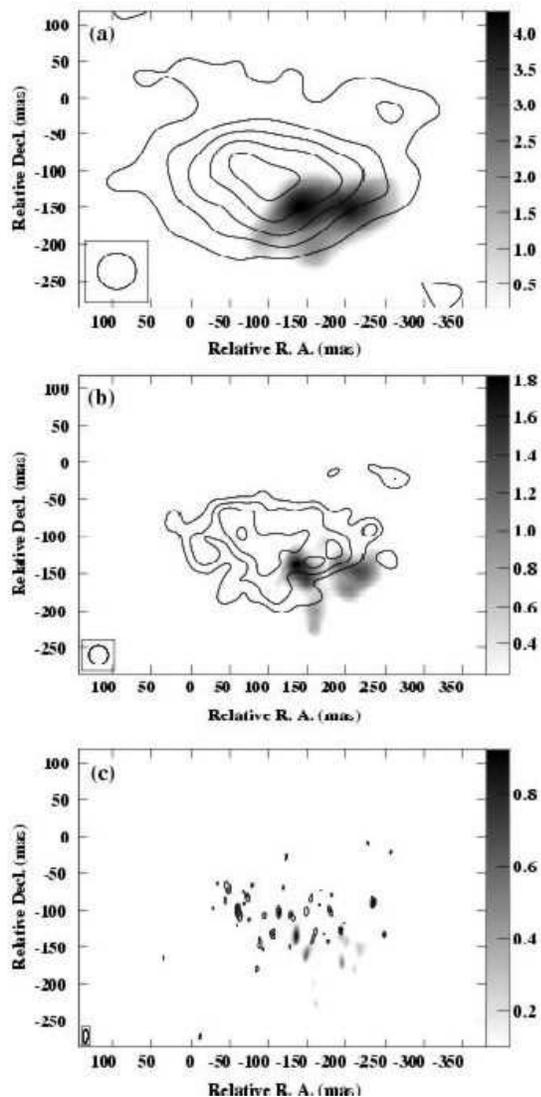}
\caption{Gray-scale total intensity (moment 0) images of the 1667~MHz OH 18~cm main line,
with 1599~MHz continuum contours superposed, at various angular resolutions. ({\it a}) At
50~mas resolution: Contour levels are at 4, 8, 12, 16, 20
$\times$ 1$\sigma$ (89~$\mu$Jy~beam$^{-1}$). ({\it b}) At 25~mas resolution: Contour
levels are at 4, 6, 8, 10 $\times$
1$\sigma$ (57~$\mu$Jy~beam$^{-1}$). ({\it c}) At the full resolution of the array
($17.6 \times 6.1$~mas, P.A.=$-4^{\circ}$):
Contour levels are at 4, 5, 6 $\times$ 1$\sigma$ (31~$\mu$Jy~beam$^{-1}$).
The reference position (0, 0)
is the same as in Fig.~1. The gray scale range in all these images is indicated by the
step wedge at the right-hand side
of each image in units of Jy~beam$^{-1}$~km~s$^{-1}$.}
\end{figure}

Fig.~5 is the velocity field (moment 1) of the 1667~MHz main line at 50~mas resolution. A
general velocity gradient is seen in the east-west direction, however, the velocity
contours seem to be distorted toward the western region. Moreover, the eastern region
itself shows a north-south velocity gradient.

\begin{figure}
\epsscale{1}
\plotone{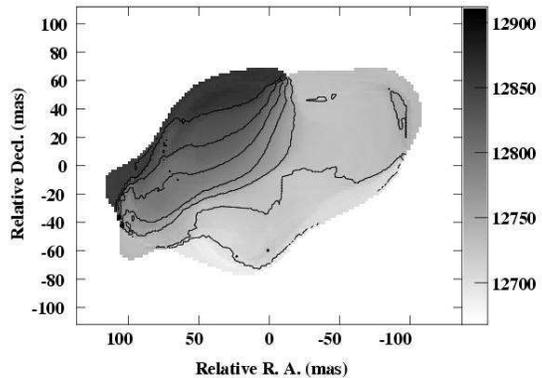}
\caption{The OH 1667~MHz velocity field (moment 1) at 50~mas resolution. Contours are from
12700 to
12850~km~s$^{-1}$ in steps of 25~km~s$^{-1}$. The gray-scale range is between 12670 and
12910 ~km~s$^{-1}$.
The reference position (0, 0) is the same as in Fig.~2.}
\end{figure}

\section{DISCUSSION}

\subsection{The Radio Continuum}

The radio continuum emission from the ULIRG IRAS~17208--0014 at 1362~MHz has been
discussed in detail by \citet{MOM03} for various angular resolutions. Our observations at
1599~MHz have imaged the radio continuum emission from IRAS~17208--0014 at a considerably
higher angular resolution than previously reported, which was $36 \times 33$~mas, P.A.
25$^{\circ}$ \citep{MOM03}.

The continuum image at the full resolution of our VLBI array (i.e., $17.6 \times 6.1$~mas;
Fig.~1) shows several compact sources with flux densities greater than 5 times the rms
noise level of 155~$\mu$Jy~beam$^{-1}$. We find no dominant compact radio source with a
very high brightness temperature. The brightness temperature of these sources are on the
order of $10^{6}$~K, indicating that the emission is non-thermal.

Following the discussion in \citet{MOM03} and references therein, these compact sources
are likely to be clustered radio supernovae (RSNe) and/or supernova remnants (SNRs), with
the likelihood that each of these sources could be mainly powered by a
recently-detonated, individual, RSN. Using the post-maximum light curve of luminous RSNe
derived by \citet{SM98}, we find that individual RSNe will lie above our 5~$\sigma$
detection level for 3.6 yr. Given the supernova rate of $4 \pm 1$~yr$^{-1}$ in
IRAS~17208--0014 \citep{MOM03}, we expect to see about $14 \pm 4$ individual RSNe. This
is in good agreement with the 10 sources revealed in our observations. 

We defer further discussion of the continuum emission to a subsequent paper, in which we
hope to report lower angular resolution results at 1599~MHz with matching-resolution
lower-frequency VLBI continuum observations at 330 and 610~MHz, and higher frequency
MERLIN continuum observations at 5~GHz.

\subsection{The OHM Emission}

The spectral-line observations reported here reveal the details of the OHM emission from
IRAS~17208--0014. The maser emission is complex, and seems to be localized in two distinct
areas, the eastern and western components, separated by 61~pc (Fig.~2). Weaker maser
emitting regions can be seen toward the north and the south.

Our VLBI observations recover $\sim$~80\% of the single dish flux density of the OHM
emission in IRAS~17208--0014 \citep{MAR89}. This suggests that extended OH emission is
resolved by even the shortest baseline in our VLBI array, namely the phased-VLA to Pie
Town baseline (52~km), and indicates that such emission regions have angular extents of
at least $0\rlap{.}{''}37$. This is well within the $\sim$~$5''$ extent of the galaxy
as seen in multi-wavelength IR observations \citep{SOI00}.

The overall nature, geometrical distribution, and kinematics of the megamaser
emission in IRAS~17208--0014 show striking similarities to those of another megamser
galaxy, III~Zw~35 \citep{PIH01,PCEP05}. These authors explain the OH emission in
III~Zw~35 by invoking the presence of an inclined clumpy ring structure oriented in a
north-south direction. A similar ring model, though with an east-west orientation, could
describe the OH emission in IRAS~17208--0014. In such a model, the two strongest emission
peaks, the eastern and western components (Fig.~2), would mark directions tangential to
the ring, and the apparent empty region in the middle would coincide with the
center of the ring. Assuming a circular ring structure, and excluding the southern and
northern regions seen in Fig.~2, the apparent extend of the OHM emission on the plane of
the sky suggests an inclination angle of $\sim55^{\circ}$. Numerical simulations and
modeling, however, are needed to derive a more reliable estimate for the inclination
angle.

In addition to the eastern and western components, the OHM emission in IRAS~17208--0014
shows a prominent extension to the south at various angular resolutions. This structure
does not seem to be part of the proposed ring structure, but lies perpendicular to it.
Considering its position, width, and velocity (Fig.~2, Region 7 and Table~2), this
feature could be associated with an inflow or an outflow. Molecular outflows and inflows
have been previously reported in several megamaser galaxies, sometimes with velocities up
to 800~km~s$^{-1}$ \citep{BHH89,PIH05}. Simulations carried out by \citet{MH96} show that
when gas-rich galaxies merge, strong gravitational torques drive the gas to the center of
the remnant as a strong inflow and fuel an intense but very short-lived starburst.
Considering that IRAS~17208--0014 is an advanced merger system powered by an extreme
starburst \citep{ZL93,MUR96,MOM03}, the southern extension could be the result of an
inflow that is fueling the starbust. Alternatively, the southern extension could be
associated with an outflow. Extreme merging conditions, and consequently powerful nuclear
activity (AGN and/or starburst), could induce superwinds in the nuclear region, and cause
material to be blown away.
However, our data cannot uniquely establish that the southern extension is an inflow or an
outflow. Considering that another, weaker, extension is seen to the north (Region 2
of Fig.~2), it might be the case that both the northern and southern extensions are part
of another rotating structure, or they may represent different maser spots in a turbulent
velocity field.

The larger extent of the continuum emission relative to that of the OHM emission in
IRAS~17208--0014 (Fig.~4{\it a}-{\it b}) indicates that the star formation in the nuclear
region of this galaxy is not confined to just around or within the proposed ring
structure, suggesting that the OH population inversion occurs only over a small region.
This is likely because the inversion of the OH main lines depends critically on the local
IR spectrum (See Parra et al.~2005, and references therein). Moreover, these two images
show that the OHM and peak continuum intensities do not coincide. This can be explained
by free-free absorption, as was suggested for III~Zw~35 \citep{PIH01}. If the optical
depth is significant at 1.6 GHz ($\tau>1$), then the emission measure, $EM = n_{e}^2l > 6
\times 10^{6}~{\rm pc~cm}^{-6}$. Assuming $l \sim 100$~mas (76~pc), then the density is
$n_{e} > 281~{\rm cm}^{-3}$, which is consistent with that expected for a dense starburst
nucleus \citep{SM96}. Our subsequent paper on multi-frequency VLBA and MERLIN continuum
observations will address the likelihood of free-free absorption in IRAS~17208--0014.
Another possibility is that this system is an incomplete merger, where the nuclei of the
progenitor galaxies have not yet completely merged. The OH ring traces one of the
original nuclei, and the star formation in the center is asymmetrically distributed
around this nucleus.

At the full resolution of the array, the lack of positional coincidence between the
continuum and megamser spots in IRAS~17208--0014 (Fig.~4{\it c}) is also found in other
OHM galaxies, e.g., Arp~220, III~Zw~35, and Mrk~273 \citep{LDSL98,PIH01,KB04}. The
derived lower limits on the maser amplification factors in IRAS~17208--0014 are up to 130.
\citet{LDSL98} and \citet{DLLS99} estimate that the amplification factors of the megamser
emission in Arp~220 and III~Zw~35 are up to 800 and 500, respectively. Furthermore, these
authors explain the lack of positional correlation between the continuum and the OHM
spots by saturated compact masing clouds. However, \citet{PCEP05}, and based on their
detailed modeling of the OHM emission in  III~Zw~35, propose instead that the masers
could be unsaturated. The brightest maser emission could occur in regions where the
continuum emission is either very weak, or below the current detection threshold, because
of the exponential effect of path length on unsaturated masers occurring in clumpy media
\citep{PCEP05}.

The measured 1667 to 1665~MHz flux density ratio at various regions of the OHM emission
varies between $3.6 \pm 0.4$ and $8.4 \pm 0.5$ (Table~2). Similar, high 1667 to 1665~MHz
ratios are reported in other OHM galaxies, such as Mrk~273 and III~Zw~35
\citep{YAT00,PIH01}. These high ratios clearly indicate that the OH molecular levels are
not in local thermodynamic equilibrium (LTE), which would yield a ratio of 1.8
\citep{HW90}. Moreover, the line ratios of the eastern and western OHM regions (Regions 1
and 5; Fig.~2 and Table~2) in IRAS~17208--0014, which are the tangents of the proposed
ring structure, are very similar to those of the tangent points in III~Zw~35
\citep{PIH01}. Based on the very high 1667 to 1665~MHz line ratios observed in Arp~220,
\citet{LDSL98} have suggested the existence of two maser emission phases. One is compact,
characterized by saturated, high gain, masers, and the other diffuse, characterized by
unsaturated, low gain, masers. However, \citet{PIH01} have proposed instead a mechanism
for OHM emission based on a single phase of OH masing small clouds ($\sim $1 pc), and
\citet{PCEP05} have argued that the observed differences in the line ratios between
compact and diffuse phases could be a natural property of one phase characterized by
clumpy unsaturated masers, in agreement with the classical megamaser model of
\citet{B89}. Furthermore, our results suggest that higher line ratios seem to be
correlated with slightly narrower 1667~MHz line widths. Regions with line ratios greater
than 6 show velocity widths between 73 and 85~km~s$^{-1}$, while regions with lower line
ratios display velocity widths between 95 and 128~km~s$^{-1}$. Therefore, higher 1667 to
1665~MHz line ratios can be explained by the exponential amplification of the background
radiation by overlapping unsaturated maser clouds in both space and velocity
\citep{PCEP05,LE05}.

The exponential amplification that causes stronger 1667~MHz lines, and consequently higher
1667 to 1665 MHz line ratios, could also explain the narrower measured widths of the
1667~MHz lines compared to those at 1665~MHz (Table~2). 
While the 1667~MHz transition exhibits a range of strong and weak features
in a given region (Figure~2), the 1665~MHz counterparts will have 
a smaller range of amplitudes, and appear as a wider, blended, spectral feature.

The similarities between the OHM emission in IRAS~17208--0014 and III~Zw~35, and
consequently the proposed ring model, are further supported by the observed velocity
field in IRAS~17208--0014 (Fig.~5). This field shows a strong gradient in the eastern
region in two perpendicular directions, east-west, and north-south, and a distorted field
in the western region. A similar velocity field has been observed in III~Zw~35
\citep{PIH01}, where the northern masing region has a two directional velocity gradient,
and the southern region has a distorted velocity field.

To explain the observed velocity field by the ring model for the OHM emission in
III~Zw~35, \citet{PCEP05} have proposed that masing clouds, in addition to their
rotational velocity, have a comparable outflow velocity parallel to the ring axis and
directed away from the ring midplane. In the terminology of these authors, a cloud near a
tangent region that lies above the ring midplane will have a projected outflow velocity
that is blueshifted, while a cloud below the ring midplane will have a redshifted
projected outflow velocity.
Such a mechanism could produce an apparent velocity difference of nearly twice the
projected outflow velocity, and would be perpendicular to the rotational velocity of the
ring. These authors further note that at the tangent point where material is moving away
from the observer, i.e., the northern tangent point in III~Zw~35, or the eastern tangent
point in IRAS~17208--0014, the cloud outflow mechanism gives a velocity gradient in the
same direction as that caused by the rotation mechanism, thus reinforcing the gradient.
In contrast, the tangent point where material is moving toward the observer, i.e., the
southern tangent point in III~Zw~35, or the western tangent point in IRAS~17208--0014,
the two mechanisms have opposite directions and will partly cancel. This could result in
the distorted velocity field seen in the southern component of III~Zw~35, or in the
western component of IRAS~17208--0014. However, \citet{PCEP05} have suggested that
alternatively the difference may mainly be due to the statistical nature of the clumpy
maser model used in their Monte-Carlo simulations.

While various regions of the OHM emission show one dominant spectral peak in the 1667~MHz
line, Region 6 (Fig.~2 and Table~2) displays a double spectral feature, suggesting the
existence of a velocity substructure in this region. Considering that bright OHM emission 
spots are the result of spatial and velocity alignment of a number of unsaturated
maser clouds \citep{PCEP05}, the observed double-peak spectrum indicates that there are
two distinct aligned groups of masing clouds within this region.

Assuming a rotational velocity of $\sim69~({\rm sin}~i)$~km~s$^{-1}$, the velocity
gradient between the eastern and western masing peaks in IRAS~17208--0014,
i.e., the tangents of the proposed ring distribution, is $\sim2.3~({\rm sin}~i)~{\rm
km~s}^{-1}~{\rm pc}^{-1}$. For a radius of 30.5~pc, we derive an enclosed dynamical mass
of about ${3 \times 10^7~({\rm sin}^{-2}~i)~M{_\odot}}$. A greater mass value is
estimated on larger scales from molecular (CO) and atomic (H~{\footnotesize I}) gas
\citep{DS98,MOM03}. Within a radius of 30.5~pc, the measured mass density from the OH
observations is $281~({\rm sin}^{-2}~i)~M{_\odot}~{\rm pc}^{-3}$.

The large scale velocity field attributed to the rotating ring structure is comparable
or smaller than the 1667~MHz line widths obtained at various regions of the ring
(Table~2). Very similar rotational velocity and velocity widths are reported in III~Zw~35
\citep{PIH01,PCEP05}. These velocity widths, although very wide compared to galactic OH
masers, seem to be typical for OHM sources. In their VLBI study of Arp~220, \citet{LDSL98}
suggest that much of the broad megamaser line width is intrinsic to the compact maser
spots. Moreover, \citet{PCEP05} explain that the observed line widths cannot arise from a
population of narrow-width clouds, but from the spectral blending of masing clouds with
internal velocity widths $\gg 1$~km~s$^{-1}$. In their detailed modeling of the OHM
emission from III~Zw~35, these authors report that in order to fit the observed velocity
width of the compact maser spots, their model required maser clouds with
internal velocity dispersion of 20~km~s$^{-1}$. The high velocity dispersion of the OHM
clouds indicates a highly turbulent medium, and suggests that the clouds are not
gravitationally confined \citep{PCEP05}.

\section{CONCLUSIONS}

We have presented the results of sensitive, phase-referenced VLBI observations of the 18
cm continuum and OH megamaser emission in the central 0.4~kpc of the ULIRG
IRAS~17208--0014.

The continuum emission from this galaxy was imaged at a higher resolution than has been
previously
reported. Several compact sources with flux densities greater than 5$\sigma$ were
detected. These sources have brightness temperatures on the order of $10^{6}$~K,
and are more likely to be clustered RSNe and SNRs.
However, we cannot rule out the possibility that each of the compact sources is mainly
powered by an individual bright RSN nested in a region that contains faded SNRs. This is
supported by the agreement in the number of the observed and expected continuum sources
above the 5~$\sigma$ level.
The continuum results do not reveal a radio-loud AGN in the nuclear region of this galaxy. 

The OHM results show the detection of both the 1665 and 1667~MHz mainlines in
IRAS~17208--0014. The OHM emission in this ULIRG has an extent of $170 \times 110$~pc,
and is mostly localized in two regions separated by 80~mas (61~pc). The overall structure
and dynamics of the maser emission seem to be consistent with a clumpy, rotating,
ring-like geometry, with the two dominant maser regions marking the tangential points of
the proposed ring structure. However, our OH results reveal another component extended to
the south. Considering the ongoing intense starburst in this advanced merger system, the
southern extension could to be associated with a gaseous inflow or outflow. 
Assuming Keplerian motion for the observed OH ring structure, the enclosed dynamical mass
and the mass density within a radius of 30.5~pc, are about {$3 \times 10^7~({\rm
sin}^{-2}i)~M{_\odot}$}, and $281~({\rm sin}^{-2}~i)~M{_\odot}~{\rm pc}^{-3}$,
respectively. 

Our results show that the starburst activity, as seen in the radio continuum images, has a
larger extent than the OHM emission. Moreover, the low-resolution images show that at
larger scales, the OHM emission peaks do not coincide with the peak of the continuum
emission. This could be the results of free-free absorption and/or possibly an indication
that the merger is incomplete. At the full angular resolution, the brightest continuum and
OHM spots are not correlated. The derived lower limits on the maser amplification are up
to 130.

The observed line-to-continuum ratios, 1667 to 1665~MHz line ratios, and the 1667~MHz line
widths suggest that the strong OHM emission in various regions can be explained by
unsaturated masing clouds overlapping in both space and velocity.

\section{ACKNOWLEDGMENT}
The authors thank C. J. Salter, J. Darling, M. Elitzur, and the anonymous referee for very
helpful comments. This research has made use of the NASA/IPAC Extragalactic Database
(NED) which is operated by the Jet Propulsion Laboratory, California Institute of
Technology, under contract with the National Aeronautics and Space Administration.

\begin{deluxetable}{ccccccc}
\tablecolumns{7}
\tablewidth{0pc}
\tablecaption{C{\footnotesize OMPACT} S{\footnotesize OURCES} {\footnotesize IN}
IRAS~17208--0014}
\tablehead{
\colhead{}    &\colhead{}    &\colhead{}
&\multicolumn{4}{c}{Gaussian Component Parameters} \\ \cline{4-7} \\
\colhead{Source} &  & \colhead{Relative Position\tablenotemark{a}}&
\colhead{Peak\tablenotemark{b}} & \colhead{Total} &
\colhead{Deconvolved Size\tablenotemark{c}} &
\colhead{P.A.} \\
\colhead{} & \colhead{} &  \colhead{(mas)} &
\colhead{($\mu$Jy~beam$^{-1}$)} & \colhead{($\mu$Jy)} & \colhead{(pc)}&
\colhead{($^{\circ}$)} \\
\colhead{(1)} & &
\colhead{(2)}& \colhead{(3)} & \colhead{(4)} & \colhead{(5)} & \colhead{(6)}}
\startdata
1\dotfill & &   0, 0     & $177  \pm 31$ & $320 \pm 81 $ & $9.3~\times 5.0   $ & 5   \\
2\dotfill & & 13W, 32S   & $196  \pm 30$ & $604 \pm 119$ & $21.3~\times 5.4  $ & 12  \\
3\dotfill & & 24W, 11S   & $164  \pm 32$ & $241 \pm 71 $ & $9.8~\times <7.1  $ & ... \\
4\dotfill & & 57W, 60S   & $168  \pm 31$ & $299 \pm 81 $ & $7.6~\times 4.4   $ & 38  \\
5\dotfill & & 65W, 31S   & $185  \pm 31$ & $295 \pm 75 $ & $10.7~\times 3.0  $ & 10  \\
6\dotfill & & 81W, 36S   & $167  \pm 31$ & $307 \pm 83 $ & $10.1~\times <8.9 $ & ... \\
7\dotfill & & 131W, 29S  & $166  \pm 32$ & $243 \pm 71 $ & $12.3~\times <2.3 $ & ... \\
8\dotfill & & 146W, 55S  & $164  \pm 32$ & $210 \pm 66 $ & $6.8~\times 2.4   $ & 7   \\
9\dotfill & & 185W, 18S  & $184  \pm 31$ & $319 \pm 79 $ & $7.1~\times 5.4   $ & 177 \\
10\dotfill & &200W, 61S  & $167  \pm 33$ & $119 \pm 45 $ & $    < 4.9        $ & ... \\
\enddata
\tablenotetext{a}{The reference position (0,0) is $\alpha\rm{(J2000)}=17^{\rm
h} 23^{\rm m} 21\rlap{.}^{\rm s} 96158$, $\delta\rm{(J2000)}=-00^{\circ} 17'
00\rlap{.}'' 8904$.}
\tablenotetext{b}{Higher than $5{\sigma}=155~{\mu}$Jy~beam$^{-1}$.}
\tablenotetext{c}{At half maximum.}
\end{deluxetable}

\begin{deluxetable}{ccccccccc}
\tablecolumns{9}
\tablewidth{0pc}
\tablecaption{G{\footnotesize AUSSIAN} P{\footnotesize ARAMETERS} {\footnotesize 
OF THE} OH M{\footnotesize EGAMASER} L{\footnotesize INES} {\footnotesize IN THE} 
R{\footnotesize EGIONS OF} F{\footnotesize IGURE} 2}\tablehead{
\colhead{} &\multicolumn{3}{c}{1667 MHz} &\colhead{} &\multicolumn{3}{c}{1665 
MHz} &\colhead{} \\\cline{2-4}  \cline{6-8} \\
\colhead{Region} &  \colhead{Velocity} & \colhead{$\Delta V_{\rm{FWHM}}$} & 
\colhead{$F_{\rm{1667}}$} &\colhead{} & \colhead{Velocity} & \colhead{$\Delta 
V_{\rm{FWHM}}$} & \colhead{$F_{\rm{1665}}$} 
&\colhead{$F_{\rm{1667}}/F_{\rm{1665}}$}\\\colhead{} & \colhead{(km s$^{-1}$)} & 
 \colhead{(km s$^{-1}$)} & \colhead{(mJy beam$^{-1}$)} &\colhead{} &\colhead{(km 
s$^{-1}$)} &  \colhead{(km s$^{-1}$)} & \colhead{(mJy beam$^{-1}$)} & \colhead{} 
\\\colhead{(1)} & \colhead{(2)}& \colhead{(3)} & \colhead{(4)} & \colhead{} 
&\colhead{(5)} & \colhead{(6)} &\colhead{(7)} & \colhead{(8)}}\startdata
1\dotfill &  $12864 \pm 2$ & $73 \pm 4$ & $12.3 \pm 0.5$  & 
&$ 13193 \pm 18$& $234 \pm 40 $ & $2.0 \pm 0.2$& $6.2 \pm 0.5$\\
2\dotfill & $12861 \pm 8$ & $97 \pm 13$ &  $4.8 \pm 0.4$  &
&$ 13217 \pm 17$& $261 \pm 44 $ & $1.3 \pm 0.2$& $3.7 \pm 0.5$\\
3\dotfill & $12727 \pm 2$ & $85 \pm 4$ &  $11.3 \pm 0.3$  &
&$ 13095 \pm 13$& $270 \pm 34 $ & $1.9 \pm 0.2$& $6.0 \pm 0.4$\\
5\dotfill & $12727 \pm 2$ & $77 \pm 4$ & $13.4 \pm 0.5$  & 
&$ 13103 \pm 15$& $263 \pm 39 $ & $1.6 \pm 0.2$& $8.4 \pm 0.5$\\
6\dotfill & $12768 \pm 2$ & $95 \pm 5$ & $8.0 \pm 0.3$   &
&$ 13140 \pm 8 $& $198 \pm 18 $ & $2.2 \pm 0.2$& $3.6 \pm 0.4$\\
          & $12875 \pm 2$ & $56 \pm 4$ & $6.1 \pm 0.4$   &
& ... & ... & ... & ... \\
7\dotfill & $12724 \pm 2$ & $128 \pm 5$ & $6.5 \pm 0.2$  &
&$ 13114 \pm 11$& $139 \pm 27 $ & $1.2 \pm 0.2$& $5.4 \pm 0.3$\\
8\dotfill & $12715 \pm 3$ & $82 \pm 5$ & $9.9 \pm 0.4$  &
& $ 13083 \pm 10$& $95 \pm 24 $&$1.2 \pm 0.3$ & $ 8.3 \pm 0.5$\\
\enddata
\end{deluxetable}

\end{document}